\documentclass[12pt]{iopart}

\usepackage{color}
\usepackage{graphicx}

\usepackage{enumerate}
\usepackage{comment}
\expandafter\let\csname equation*\endcsname\relax
\expandafter\let\csname endequation*\endcsname\relax
\usepackage{amsmath}
\usepackage{amssymb}
\usepackage[dvipsnames]{xcolor}
\usepackage{tikz}
\usepackage{amsthm}
\usepackage{pict2e}
\usepackage{bbold}
\usepackage{empheq}
\usepackage{subfigure}
\usepackage{xfrac}
\usepackage{hyperref}
\usepackage{enumitem}
\usepackage{notes2bib}
\usepackage[numbers]{natbib}
\usepackage{float}
\usepackage{natbib}

\begin{document}

\title[Complex dynamics in circular and deformed billiards with anisotropy and strain]{Complex dynamics in circular and deformed bilayer graphene inspired billiards with anisotropy and strain}

\author{Lukas Seemann}
\address{Institute of Physics, Technische Universit\"at Chemnitz, D-09107 Chemnitz, Germany}
\ead{lukas.seemann@physik.tu-chemnitz.de}

\author{Jana Lukin}
\address{Institute of Physics, Technische Universit\"at Chemnitz, D-09107 Chemnitz, Germany}

\author{Max H\"a\ss ler}
\address{Institute of Physics, Technische Universit\"at Chemnitz, D-09107 Chemnitz, Germany}

\author{Sibylle Gemming}
\address{Institute of Physics, Technische Universit\"at Chemnitz, D-09107 Chemnitz, Germany}

\author{Martina Hentschel}
\address{Institute of Physics, Technische Universit\"at Chemnitz, D-09107 Chemnitz, Germany}

\vspace{10pt}
\begin{indented}
\item[]December 10, 2024
\end{indented}

\begin{abstract}
While billiard systems of various shapes have been used as paradigmatic model systems in the fields of nonlinear dynamics and quantum chaos, few studies have investigated anisotropic billiards. Motivated by the tremendous advances in using and controlling electronic and optical mesoscopic systems with bilayer graphene representing an easily accessible anisotropic material for electrons when trigonal warping is present, we investigate billiards of various anisotropies and geometries using a trajectory tracing  approach founded in the concept of ray-wave correspondence. We find that the presence of anisotropy can render the billiards' dynamics dramatically from its isotropic counterpart. It may induce chaotic and mixed dynamics in otherwise integrable systems, and may stabilize originally unstable trajectories. We characterize the dynamics of anisotropic billiards in real and phase space using Lyapunov exponents and the Poincaré surface of section as phase space representation.

\end{abstract}

\newpage
\section{Introduction}
Billiards of all kinds have proven to be versatile model systems of mathematics and physics. Besides a variation of the shape, or geometry, of hard-wall billiards typically studied in the field of (classical) nonlinear dynamics, one can use physical inspiration and change their properties such as boundary conditions or equations of motion to apply to electrons or light \cite{Akkermans}. This has been done in the context of quantum chaos \cite{Stoeckmann} for example for electronic billiards realized in the form of quantum dots or light confined in dielectric cavities of certain shapes by total internal reflection. An important topic has been the investigation of wave billiards and their classical counterparts, where a wave-ray, or quantum-classical correspondence is expected and was confirmed in many examples, e.g. \cite{chaotic_light_Stone_nature, annbill}.

Here, we will focus on anisotropic billiards for electrons, 
inspired by the technological progress in the fabrication of bilayer graphene (BLG) samples {\cite{Nyakiti_NanoLett_2012, Pakdehi_ANM_2019, Ciuk_ApplSurfSci_2024} employing their unique properties. Epitaxial graphene even finds applications in metrology, e.g. as reference for Kelvin probe force microscopy \cite{Ciuk_ApplSurfSci_2024}. Here, we focus on Bernal-stacked BLG (BBLG), which has very recently become a topic of in-depth experimental and theoretical investigations: A plethora of ordered electronic phases have been obtained in external fields \cite{SeilerNatComm_2024}, which comprise spin and valley magnetism, correlated insulators, correlated metals, charge and spin density waves and superconductivity \cite{SeilerPRL_2024,Dong_PRB_2024,Dong_arXiv_2024}.

 In the absence of external fields, neutral BBLG is semiconducting as a result of the interlayer interaction of the two graphene sheets; applying a transverse electric field lifts the equivalence of the two layers (and may induce a so-called gap) and at low temperatures the resistance varies non-monotonously with the gap closing and reopening with increasing field strength \cite{Weitz_Sci_2010,Bao_PNAS_2012,Velasco_NatNano_2012}. A DFT-based model Hamitonian recently attributed this finding to a charge-ordered state, which originates from an interplay of interlayer van der Waals interactions and ripple formation \cite{Jiang_PRR_2024}. 

 Adding extrinsic charge carriers shifts the Fermi level into the frontier states at the $K$ and $K'$ points (see below), where the threefold symmetry of the graphene lattice and the resulting trigonal warping are reflected in the trigonal symmetry of the resulting Fermi lines \cite{SeilerNatComm_2024,Koh_arxiv_2024}. With increasing carrier density the Fermi line evolves from a small, all convex, onigiri-type (inspired by the Japanese rice snack of similar geometry) triangle with rounded corners to a larger triangle with concave edges. Adding a transverse electric field enhances the concave inward bending of the edges, which is predominantly observed for holes \cite{SeilerNatComm_2024}. Similar transitions from convex to concave triangular shapes have also been obtained theoretically with model Hamiltonians, which include coupled charge and spin density waves \cite{Dong_arXiv_2024}. For weak Hund's coupling in addition to long-range Coulomb and with Ising-type spin-orbit coupling predominantly concave trigonally symmetric shapes dominate \cite{Koh_arxiv_2024}.

The manifold all-electrical control mechanism for BLG via gate voltages and their importance for potential applications based on electronic transport motivate the studies presented here. The anisotropy we first focus on (chapter \ref{chap2}) is the preference for   
certain propagation directions in space in the form of trigonal warping at the K-points. 
As outlined above, in BBLG this can be obtained by proximity with a spin-orbit-coupled two-dimensional (2D) layer \cite{Koh_arxiv_2024}, by strong doping with charge carriers \cite{SeilerPRL_2024}, or by applying external fields \cite{Dong_PRB_2024,Dong_arXiv_2024}. According to the theorem of Emmy Noether, angular momentum then cannot be conserved anymore -- and that is the moment when symmetry considerations are in order. Indeed, the dynamics of the anisotropic billiard has to differ from its isotropic counterpart. In fact, chaotic dynamics was observed in circular BLG billiards \cite{SeemannKnHe_BLGI_PRB} as well as in BLG billiards of noncircular shapes \cite{Seemann_Knothe_Hentschel_2024}. 

Notice that there are tow types of $K$-points (so-called valleys) with mirrored Fermi lines in BLG. Since we do not consider intervalley scattering and want to keep the system as simple as possible, we focus on one $K$-point throughout. This scenario corresponds to isospin-conserving electron dynamics within one valley, which may be obtained in Bernal-stacked BLG at low carrier densities, i.e. in the weakly coupling regime and for external fields, which lift the degeneracy of the multitude of valleys. We will discuss the consequences of the electron's pseudospin on the billiards dynamics in detail below, but point out that it breaks the mirror symmetry of the Poincaré surface of section (PSOS) for a trigonally-warped Fermi line, see the right panel of Fig.~\ref{fig:chap2_1}, while a deformed billiard shape with isotropic dispersion always possesses this symmetry, cf.~left panel of Fig.~\ref{fig:chap2_1}.

Mechanical strain is a further source of anisotropy, here induced by an applied external force \cite{Liu_arxiv_2024}, and we will consider this in the second part of the paper in Chapter \ref{chapstrain}. We will focus on closed (hard wall) billiards in this systematic ray-tracing study. We close with a summary and outlook and argue that based on ray-wave correspondence our results are relevant for realistic (quantum) systems.

\begin{figure}
    \centering
    \includegraphics[width=1\textwidth]{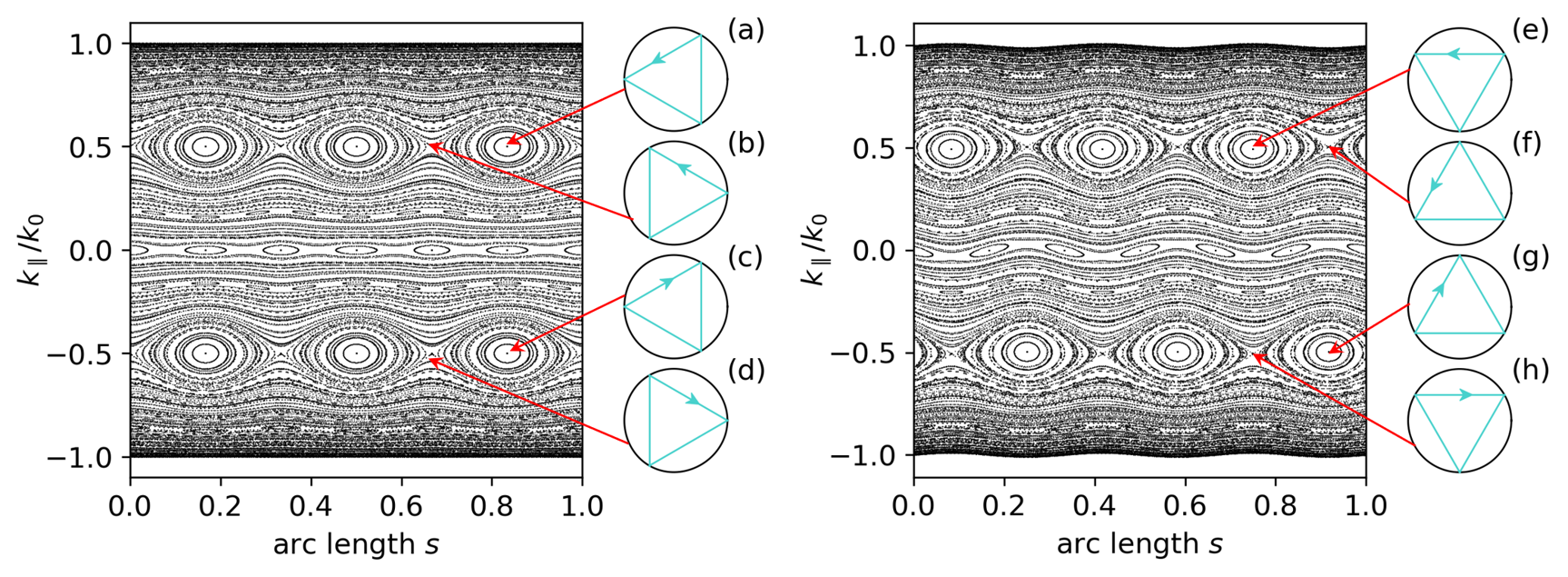}
    \caption{Stable and unstable triangular orbits in an onigiri shaped cavity ($\epsilon_{3g}=0.01$) with
    an circular Fermi line on the left and a circular cavity with
    an onigiri shaped Fermi line ($\epsilon_{3f}=0.01$) on the right. The stable orbits (a), (c), (e) and (g) correspond to
    elliptical fixed points in the PSOS, while the unstable orbits (b), (d), (f) and (h) correspond to
    hyperbolic fixed points.}
    \label{fig:chap2_1}
\end{figure}

\section{Circular and deformed billiards in anisotropic materials}
\label{chap2}
We investigate the billiards dynamics in real and phase space using a ray-tracing algorithm. Therefore we need a link between the anisotropic momentum space and the real space billiards dynamics. This link is provided by the group velocity (or the Poynting vector) which is always normal to the Fermi line (or to the index ellipsoid for optical systems). Thus, a certain wave vector $\vec{k}$ leads to a group velocity, which is, in anisotropic media, usually not parallel to the wave vector.
The angles of incidence $\alpha$ and reflection $\beta$ at the billiard boundary  
are derived from the conservation of the momentum component $k_\parallel$ parallel to the boundary interface and are given as the angle to the corresponding group velocity, cf. Fig.\ref{fig:angle_in_re}. 
For an isotropic, thus circular dispersion relation (Fermi line), we immediately recover the well-known reflection law $\alpha=\beta$. This relation even holds for situations where the Fermi line is symmetric with regard to the $k_\parallel$ axis. However, for generically anisotropic media such a situation is usually not present, and consequently the angle of incidence is not equal to the angle of reflection. Consequently, the symmetry breaking in the momentum space destroys the so-called whispering-gallery orbits characteristic for circular cavities and replaces them by a complex dynamics with chaotic features even in disk resonators.
\begin{figure}
    \centering
    \includegraphics[width=0.4\linewidth]{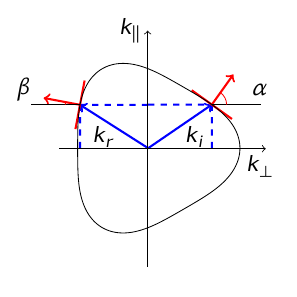}
    \caption{Angle of incidence and reflection in anisotropic media. The incident $\vec{k}_\mathrm{in}$ and reflected $\vec{k}_\mathrm{r}$ wave vectors follow from the conservation of the parallel component $k_\parallel$ of the wave vector. The normal vectors on the Fermi line, which are parallel to the group velocity, give the angles of incidence $\alpha$ and reflection $\beta$. Since the Fermi line is not symmetric to the $k_\parallel$ axis, the angle of incidence is not equal to the angle of reflection \cite{SeemannKnHe_BLGI_PRB}.}
    \label{fig:angle_in_re}
\end{figure}
\subsection{Lyapunov exponents as characteristics of the billiard's dynamics}

In this section we investigate the nonlinear dynamics in billiards of various shapes while simultaneously changing their dispersion relation. Motivated by the trigonal warping of the Fermi line in BLG, we focus on this type of geometry both in real and momentum space and refer to it as onigiri shape. 
The geometry of the cavity \textit{in real space }is defined in polar coordinates as follows
\begin{equation}
   R(\phi) = R_0(1+\epsilon_{ng} \cos (n(\phi-\theta_0)) \label{eq:ch2_3}
\end{equation}
where $R_0$ is the mean radius of the cavity and $\epsilon_{ng}$ is a deformation parameter. For $n=0$ a circle is obtained and for $n=3$ the $C_3$-symmetric onigiri geometry.

The geometry of the Fermi line \textit{in momentum space} is defined analogously to that of the cavity: 
\begin{equation}
   R_f(\phi+\theta_0) = k_{0}(1+\epsilon_{nf} \cos (n\phi)) \label{eq:ch2_4}
\end{equation}
The rotation angle $\theta_0$ indicates the tilt between the cavity geometry axis and the underlying BLG lattice 
or the Fermi line that becomes important 
if both the cavity and the Fermi line are noncircular. Here, $\theta_0$ is defined such that the geometry is rotated counter clockwise by this angle 
with respect to the (fixed) BLG lattice or Fermi line.

In a previous work \cite{Seemann_Knothe_Hentschel_2024} adressing realistic BLG systems it was found that a highly structured PSOS  
-- concerning the presence of islands and their size -- was 
achieved when the shapes of the Fermi line and and the billiard cavity possessed a very similar geometry. Then, the Poincaré SOS of the billiards showed a high portion of regular trajectory dynamics confined to, correspondingly, large islands. Moreover, these emerging island chains protected whispering gallery (WG) trajectories by hindering ("blocking") their transition into the chaotic sea of the Poincaré SOS. It is desirable to quantify these qualitative statements to allow for a profound comparison and optimization of the different geometries. To this end we introduce generalized Lyapunov exponents in order to characterize the chaotic behavior of the overall billiards dynamics.

The Lyapunov exponent is used to make statements about the stability of a given trajectory. In order to determine the Lyapunov exponent of the reference trajectory, a trajectory infinitesimally adjacent to it is considered. Let the perturbation $\epsilon(t)$ be the distance between the two trajectories at any point in time. For $t \rightarrow \infty$ and $\epsilon(t_0) \rightarrow 0$, the Lyapunov exponent $\lambda$ is defined as the mean exponential divergence or convergence of the trajectories \cite{WOLF1985285, RevModPhys.57.617},
\begin{equation}
    \epsilon (t) \approx  e^{\lambda t} \cdot \epsilon (t_0) \:,\label{eq:ch2_1}
\end{equation}
\begin{equation}
   \lambda = \frac{1}{t} \cdot \ln{\frac{\epsilon(t)}{\epsilon (t_0)}} \:.\label{eq:ch2_2}
\end{equation}
The perturbation $\epsilon(t)$ is limited from above, since it cannot be larger than half the circumference of the cavity. Thus it is necessary to renormalize the perturbation to $\epsilon(t_0)$ when it becomes larger than an upper limit $\epsilon_\mathrm{max}$ and to calculate the Lyapunov exponent step by step from renormalization to renormalization or until the end of the trajectory \cite{WOLF1985285}.

A positive Lyapunov exponent characterizes exponentially drifting apart trajectories. The reference trajectory is then chaotic, i.e. unstable. A stable trajectory, on the other hand, has a Lyapunov exponent that is equal to zero. The choice of the reference trajectory is therefore not arbitrary in a mixed phase space. In order to obtain a characteristic value for the entire system, the mean value of the Lyapunov exponents of the trajectories must be calculated.

\subsection{Deviation from the circular geometry in real and momentum space}

\begin{figure}
    \centering
    \includegraphics[width=0.5\textwidth]{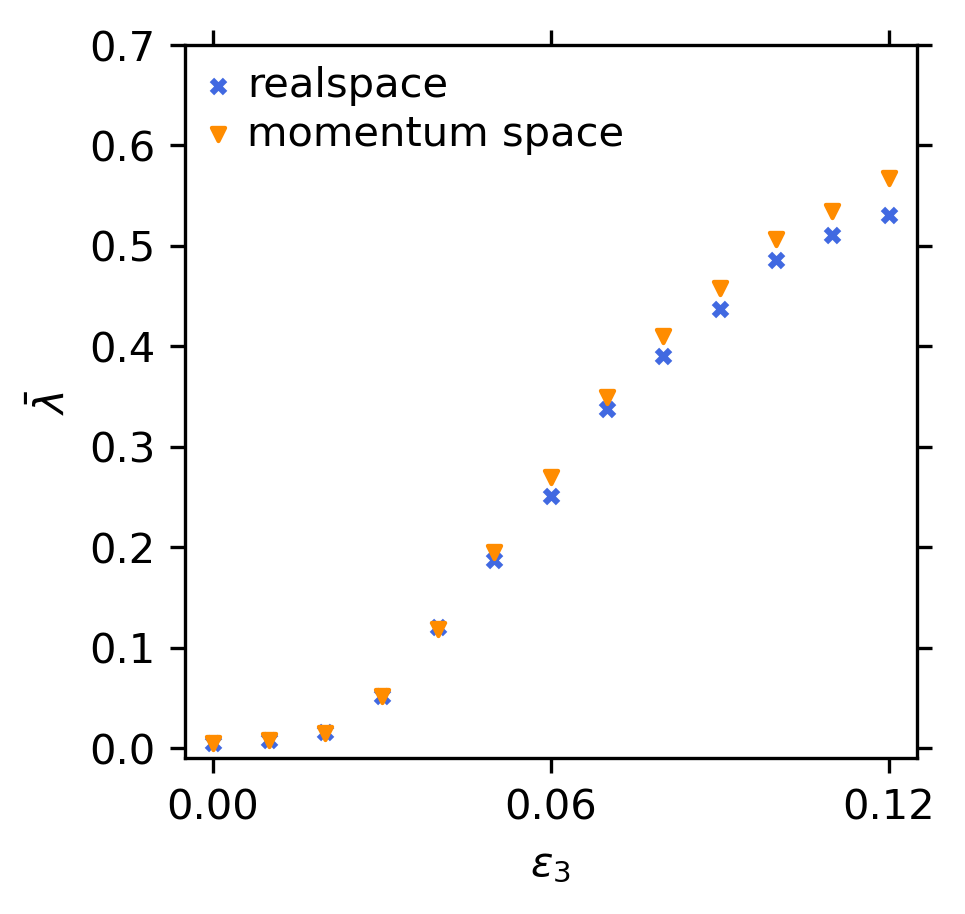}
    \caption{Average Lyapunov exponent $\bar{\lambda}$ as a function of the deformation parameter $\epsilon_3$. The circular symmetry was broken in real space or momentum space by varying the deformation parameter of the cavity or the Fermi line. A larger $\epsilon_3$ leads to a larger $\bar{\lambda}$. A deformed Fermi line results in a larger Lyapunov exponent than a deformed geometry.}
    \label{fig:chap2_lja1}
\end{figure}
 
We start our analysis by using the average Lyapunov exponent to investigate whether breaking the circular symmetry in momentum space or in real space has a greater influence on the cavity dynamics. 
To this end, the deformation parameter $\epsilon_3$ of the Fermi line ($\epsilon_{3f}$) or the cavity ($\epsilon_{3g}$) was varied from 0 to 0.12, while the cavity or the Fermi line remained circular. For each cavity-Fermi line pair, the mean Lyapunov exponent $\bar{\lambda}$ and its error were calculated and plotted as a function of the deformation parameter $\epsilon_3$ in Fig.~\ref{fig:chap2_lja1}.

It can be seen in Fig.~\ref{fig:chap2_lja1} that a larger deformation parameter both in real and momentum space leads to a larger $\bar{\lambda}$. This is due to the fact that chaotic behavior increasingly dominates. Furthermore, it can be seen that for higher deformation parameters a deformed Fermi line results in a larger Lyapunov exponent than a deformed geometry. Consequently, the deformation of the Fermi line has a greater influence on the dynamics of the electrons. One possible reason for this is that when a trajectory moves along the wall of the cavity, the deviation from the circular curvature is very small for each reflection. The curvature of the Fermi line, on the other hand, plays a role in every reflection, as the normal vector indicates the direction of the electrons.

\begin{figure}
    \centering
    \includegraphics[width=1\textwidth]{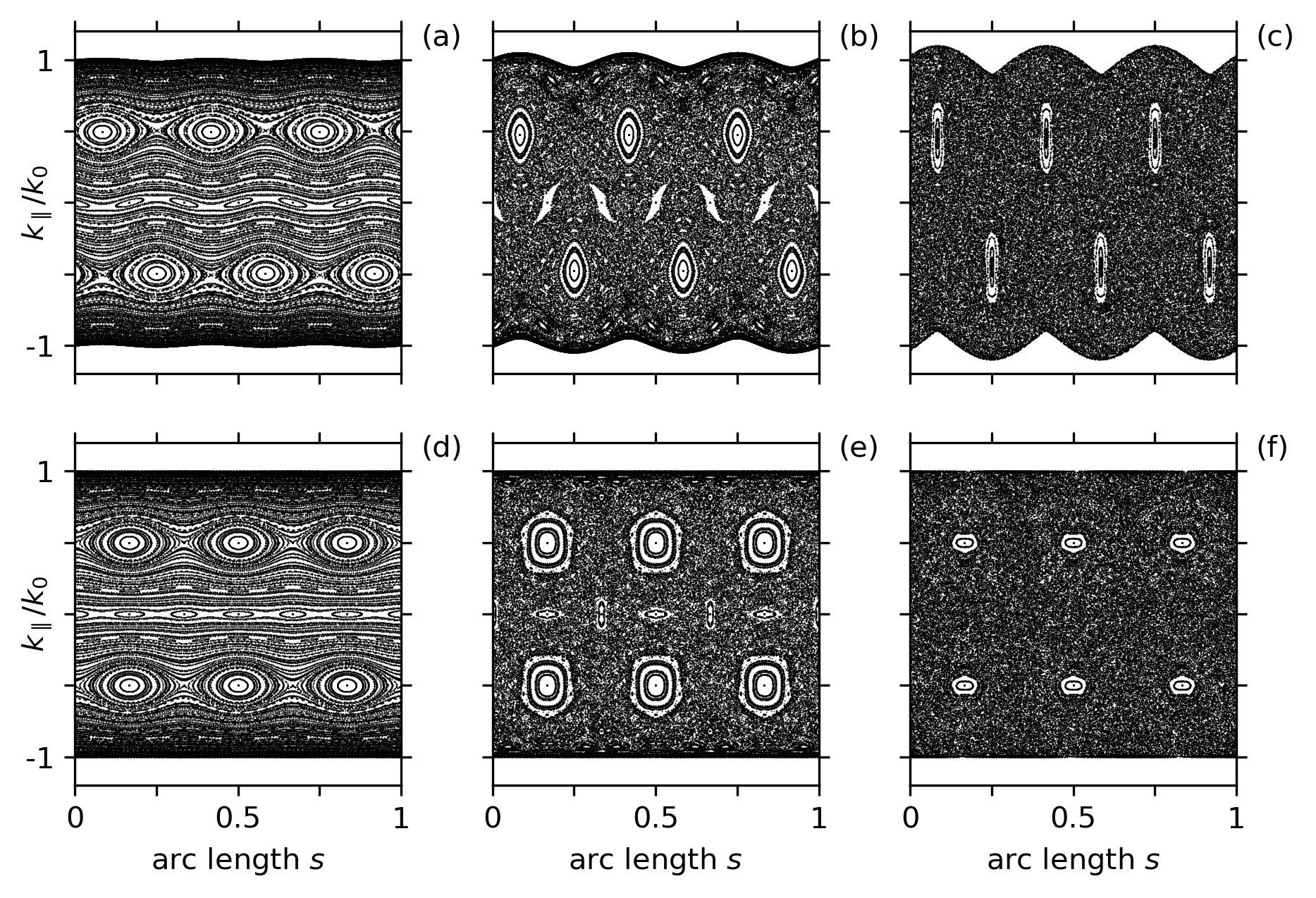}
    \caption{ (a) - (c) PSOS for circular cavities in real space and onigiri shaped Fermi lines in momentum space. (a) $\epsilon_{3f}=0.01$, (b) $\epsilon_{3f}=0.05$, (c) $\epsilon_{3f}=0.10$. The main characteristics are six islands corresponding to triangular orbits. Notice the shift between clockwise (cw) and counterclockwise (ccw) sense of rotation. 
    The larger $\epsilon_{3f}$, the more chaotic the movement. For example, the six islands, which belong to the two stable triangular orbits, become smaller.
    (d) - (f) PSOS of onigiri shaped isotropic cavities. (d) $\epsilon_{3g}=0.01$, (e) $\epsilon_{3g}=0.05$, (f) $\epsilon_{3g}=0.10$. Here, the angle of incidence and the angle of reflection are always equal and therefore the PSOS is symmetric with respect to the $k_\parallel=0$ axis. }
    \label{fig:chap2_symmetriebr}
\end{figure}

Figure \ref{fig:chap2_symmetriebr} (a)-(c) shows the Poincar\'{e} sections of a circular cavity with an onigiri shaped Fermi line with different deformation parameters $\epsilon_{3f}$. The PSOS for the reversed case of an onigiri shaped cavity with an circular Fermi line are shown in Fig.~\ref{fig:chap2_symmetriebr} (d)-(f). Here, an interpretation in terms of the Kolmogorov-Arnold-Moser (KAM) theorem \cite{KAM-Chierchia:2010}is in order. 

In the case of a small deformation parameters the PSOS is mostly filled with wavy lines, which can be seen in Fig.~\ref{fig:chap2_symmetriebr} (a) and (d). They represent deformed KAM surfaces, 
while the share of chaotic parts increases with the deformation parameter. Furthermore, in both cases one can see six big islands, which belong to two stable triangular orbits. In the case of a deformed cavity, these two orbits are stabilized by the three flat sides of the onigiri geometry. On the other hand, a deformed Fermi line results in an anisotropic velocity distribution with, in the case of the onigiri geometry, three preferred propagation directions. 
These three preferred directions create two stable triangular orbits leading to the six big islands in Fig.~\ref{fig:chap2_symmetriebr}(a)-(c). However, the upper and the lower row of these islands are not symmetric to the $k_\parallel=0$ axis (in contrast to Fig.~\ref{fig:chap2_symmetriebr}(d)-(f)) but shifted by $\Delta s=\frac{1}{6}$. In fact, there are also triangular orbits in between the islands but these are unstable, since their group velocities correspond to the curved sides of the Fermi line. 

In both cases, deformed Fermi line or cavity, the islands shrink by increasing the deformation parameter $\epsilon_3$. In the next section we will discuss, how the interplay of cavity and Fermi line deformation can increase or decrease the islands' size and whether there is an optimal setting for the largest possible islands.

\subsection{Influence of the interplay between the deformation of the cavity and Fermi line}
\begin{figure}
    \centering
    \includegraphics[width=0.5\textwidth]{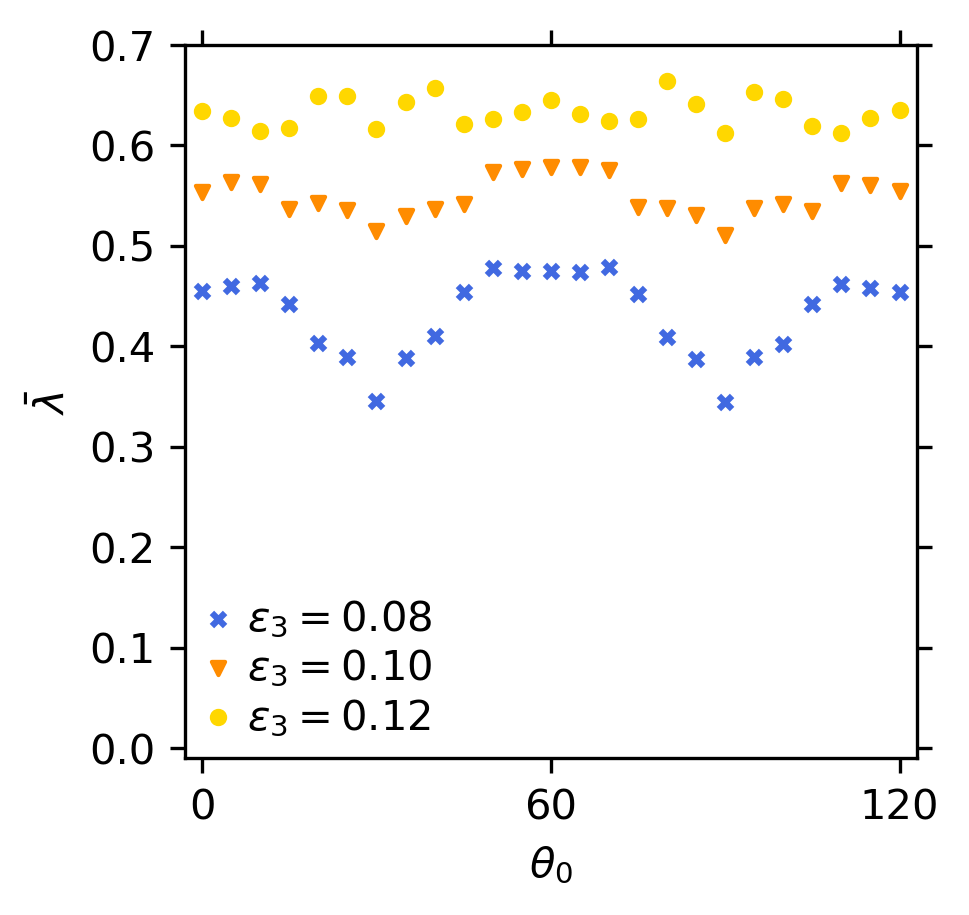}
    \caption{Average Lyapunov exponent $\bar{\lambda}$ as a function of the orientation angle $\theta_0$ for three different o'nigiri-shaped cavities with onigiri-shaped Fermi lines ($\epsilon_{3g}=\epsilon_{3f}=\epsilon_3$). A larger $\epsilon_3$ results in a larger $\bar{\lambda}$. A minimum occurs at $\theta_0=30^\circ$ and $90^\circ$.}
    \label{fig:chap2_lja2}
\end{figure}

\begin{figure}
    \centering
    \includegraphics[width=1\textwidth]{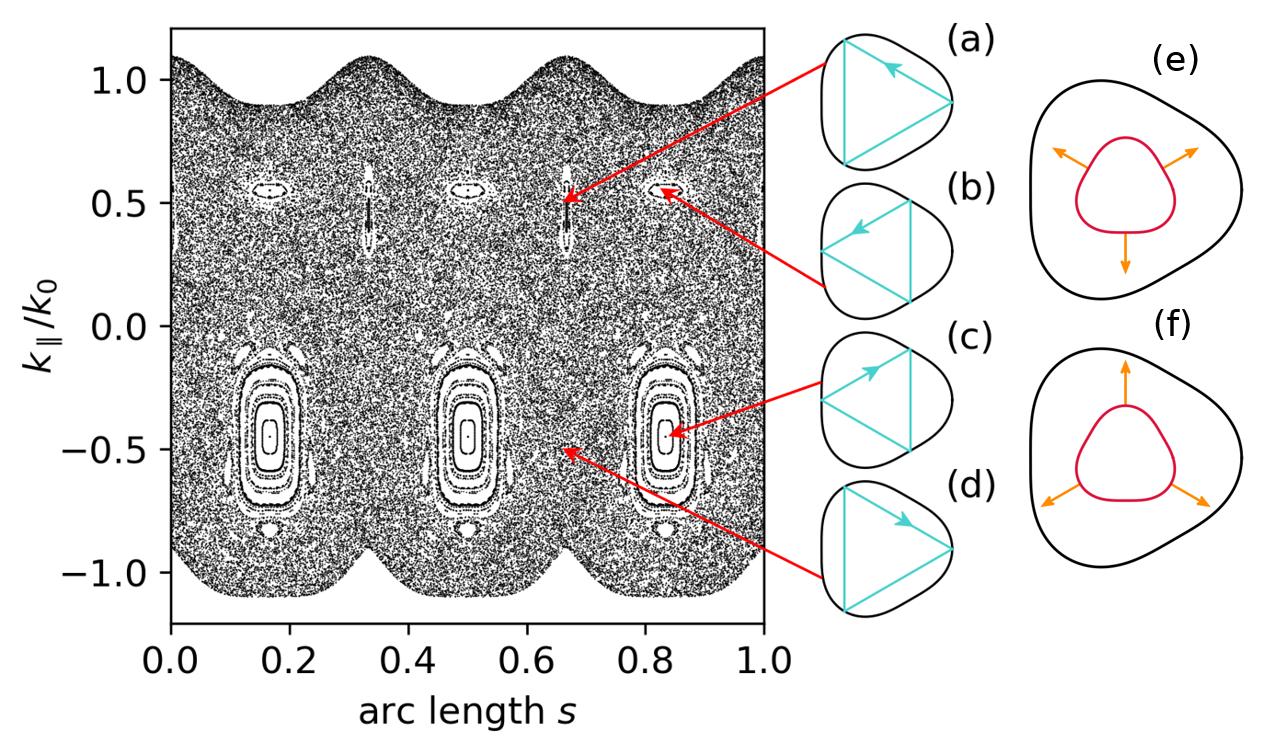}
    \caption{PSOS for an onigiri shaped cavity and Fermi line ($\varepsilon_{3g}=\varepsilon_{3f}=0.1$) at an orientation angle of $\theta_0=30^\circ$. The triangular orbits are stabilized by the (a) cavity, (b) Fermi line, (c) cavity and Fermi line, (d) neither. (e) The preferred directions (orange) of the Fermi line (red) are the propagation directions of the trajectories in (a) and (c). (f) Propagation directions of the triangular orbits in (b) and (d).}
    \label{fig:chap2_thetavar30}
\end{figure}

In this section we deform both, the Fermi line and the cavity. Then, the orientation angle $\theta_0$ between the Fermi line and the cavity becomes crucial. As long as at least one of both is circularly symmetric, a varied  orientation $\theta_0$ only results in a shift in phase space. However, now the tilt angle $\theta_0$ affects the whole dynamics of the system.

In Fig.~\ref{fig:chap2_lja2} the Lyapunov exponents for equally shaped cavities and Fermi lines, i.e. $\varepsilon_{3f}=\varepsilon_{3g}=\varepsilon_3$, are shown depending on the orientation angle $\theta_0$ (we present only the interval $\theta_0\in[0,120^\circ]$ corresponding to the C$_{\mathrm{3}}$ symmetry of the Fermi line and the cavity). In all cases, and most obvious for $\varepsilon_3=0.08$, there are minima at $\theta_0=30^\circ$ and $\theta_0=90^\circ$. These minima are due to the interplay of the stabilization properties in the momentum and the real space. Fig.~\ref{fig:chap2_thetavar30} explains this interplay. Fig.~\ref{fig:chap2_thetavar30}(a) shows a triangular orbit, where the preferred directions (corresponding to the flat sides of the Fermi line, see Fig.~\ref{fig:chap2_thetavar30}(e)) points to the regions of the cavity with high curvature, which have a destabilizing effect. Thus the islands in the PSOS become a little bit smaller. However the unstable directions, cf. Fig.~\ref{fig:chap2_thetavar30}(f), points to the stabilizing flat sides of the cavity so they become stable and form additional islands in the PSOS, see Fig.~\ref{fig:chap2_thetavar30}(b). If both stabilizing effects coincide huge islands arise, cf. Fig.~\ref{fig:chap2_thetavar30}(d), which reduce the Lyapunov exponent. In addition, the arising chain of six islands in the upper half established a barrier in phase space that protects the WG-like trajectories from entering the chaotic sea, further reducing the Lyapunov exponent.

In Fig.~\ref{fig:chap2_fermivar}(a) we show the Lyapunov exponents for different orientation angles $\theta_0$ of a system with onigiri shaped Fermi line and cavity over the Fermi line deformation parameter $\varepsilon_{3f}$, while the deformation parameter of the cavity $\varepsilon_{3g}$ is constant. The Lyapunov exponent reaches a clear minimum if the orientation angle is $\theta_0=30^\circ$ and the deformation parameter of the cavity is equal to the one of the Fermi line. Fig.~\ref{fig:chap2_fermivar}(c) shows how optimal settings for  $\varepsilon_3$ and $\theta_0$ can be used to maximally structure the phase space via the presence of large stable islands.  
However, further increasing of the deformation parameter leads to higher Lyapunov exponents and the trend of the dependency is similar to that shown in Fig.~\ref{fig:chap2_lja1}, i.e. the symmetry breaking in real space has barely any influence on the Lyapunov exponent once the anisotropy is strong enough.  

In Fig.~\ref{fig:chap2_geovar}(a) the cavity deformation parameter $\varepsilon_{3g}$ was varied while the Fermi line deformation $\varepsilon_{3g}=0.06$ remains constant. Again the average Lyapunov exponent is minimal when the orientation angle is $\theta_0=30^\circ$ and both deformation parameters are equal. 

\begin{figure}
    \centering
    \includegraphics[width=1\textwidth]{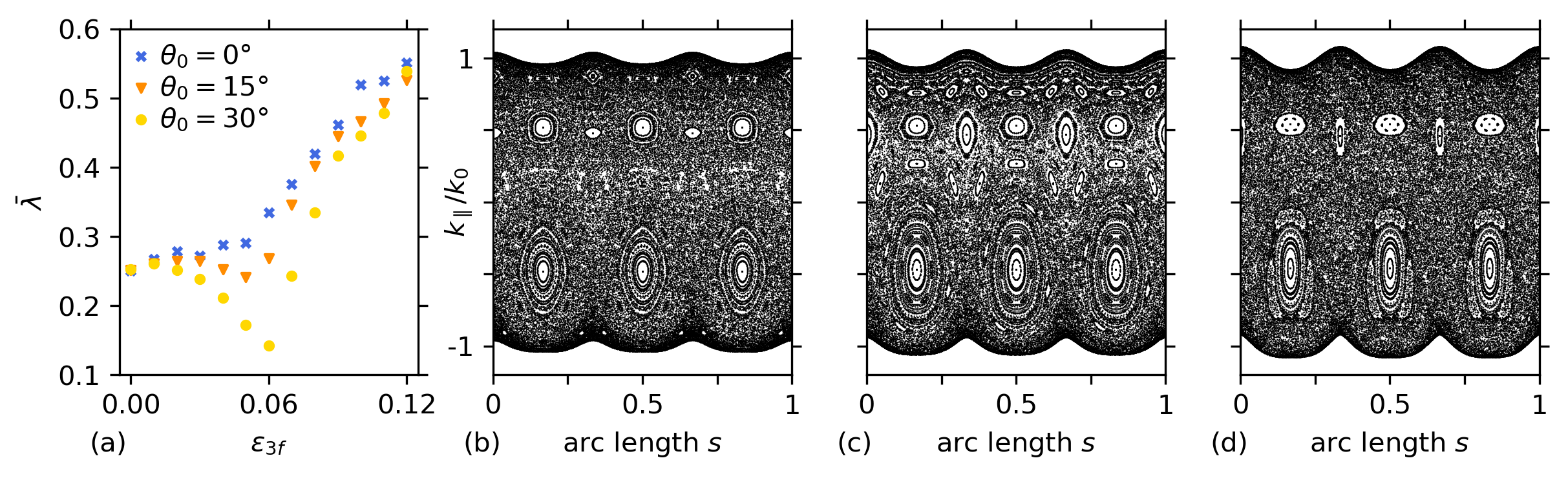}
    \caption{(a) Average Lyapunov exponent $\bar{\lambda}$ as a function of the deformation parameter $\epsilon_{3f}$ at three different orientation angles $\theta_0$. $\epsilon_{3g}$ is left constant at 0.06. A minimum occurs where the two deformation parameters $\epsilon_{3g}=\epsilon_{3f}=0.06$ coincide, but only at $\theta_0=30^\circ$.
    (b)-(d) PSOS of three onigiri shaped cavity-Fermi line pairs at an orientation angle of $\theta_0=30^\circ$. The cavity has the deformation parameter $\epsilon_{3g}=0.06$, the deformation parameter of the Fermi line is (b) $\epsilon_{3f}=0.04$, (c) $\epsilon_{3f}=0.06$, (d) $\epsilon_{3f}=0.08$. If $\epsilon_{3g}$ and $\epsilon_{3f}$ coincide, the proportion of stable orbits is the largest.}
    \label{fig:chap2_fermivar}
\end{figure}

\begin{figure}
    \centering
    \includegraphics[width=1\textwidth]{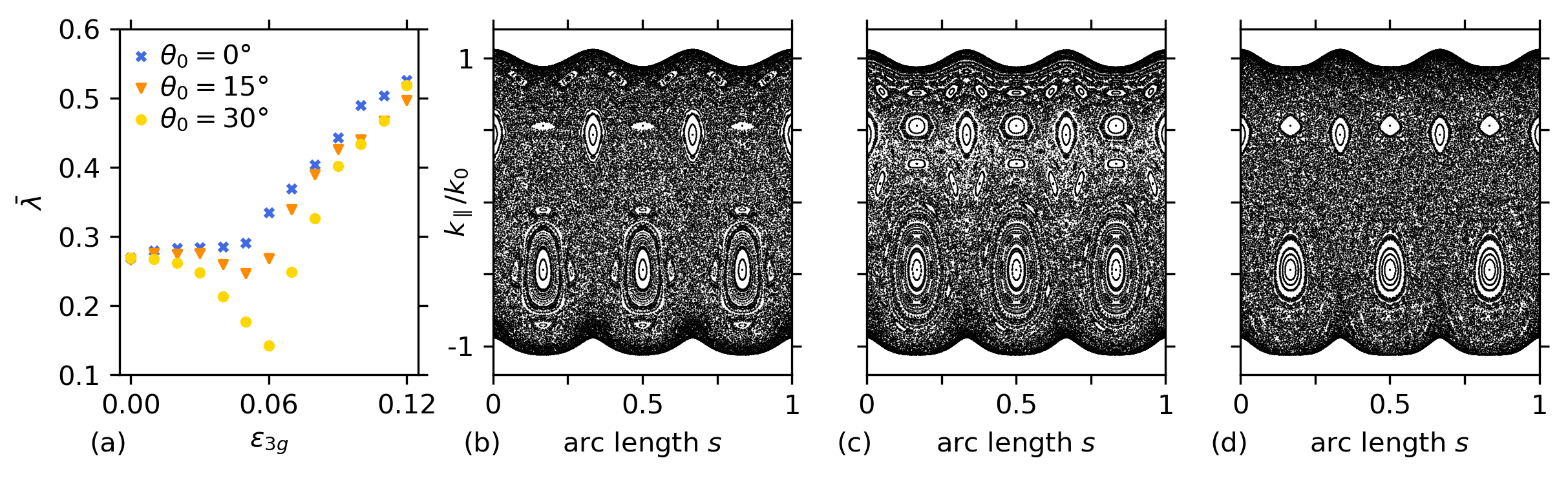}
    \caption{(a) Average Lyapunov exponent $\bar{\lambda}$ as a function of the deformation parameter $\epsilon_{3g}$ at three different orientation angles $\theta_0$. $\epsilon_{3f}$ is left constant at 0.06. A minimum occurs where the two deformation parameters $\epsilon_{3g}=\epsilon_{3f}=0.06$ coincide, but only at $\theta_0=30^\circ$.
    The Fermi line has the deformation parameter $\epsilon_{3f}=0.06$, the deformation parameter of the cavity is (b) $\epsilon_{3g}=0.04$, (c) $\epsilon_{3g}=0.06$, (d) $\epsilon_{3g}=0.08$. If $\epsilon_{3g}$ and $\epsilon_{3f}$ coincide, the proportion of stable orbits is the largest.}
    \label{fig:chap2_geovar}
\end{figure}

In order to further examine the influence of the deformation on the Lyapunov exponent, we chose a fixed orientation angle $\theta_0=30^\circ$ and vary $\varepsilon_{3g}$ for several constant $\varepsilon_{3f}$ and vice versa in Fig.~\ref{fig:chap2_lja3}. Starting with $\varepsilon_{3g}=0$ ($\varepsilon_{3f}=0$) a larger $\varepsilon_{3f}$ ($\varepsilon_{3g}$) leads to higher Lyapunov exponents, since stronger deformation of the Fermi line (cavity) results in a larger amount of chaotic trajectories if the cavity (Fermi line) is not deformed. However the Lyapunov exponent decreases by increasing the respective other deformation parameter until both are equal. 
We find clear minima at $\varepsilon_{3f}=\varepsilon_{3g}$ if $\varepsilon_{3f}<0.07$ or $\varepsilon_{3g}<0.07$. Increasing the deformation parameter $\varepsilon_{3g}$ ($\varepsilon_{3f}$) further yielding higher Lyapunov exponents but the differences between the several $\varepsilon_{3f}$ ($\varepsilon_{3g}$) are less striking. We conclude that the deformation of the Fermi line (cavity) does not play a role if the deformation of the cavity (Fermi line) is much stronger, a behavior similar to the one described in  
Fig.~\ref{fig:chap2_lja1}.
\begin{figure}
    \centering
    \includegraphics[width=1\textwidth]{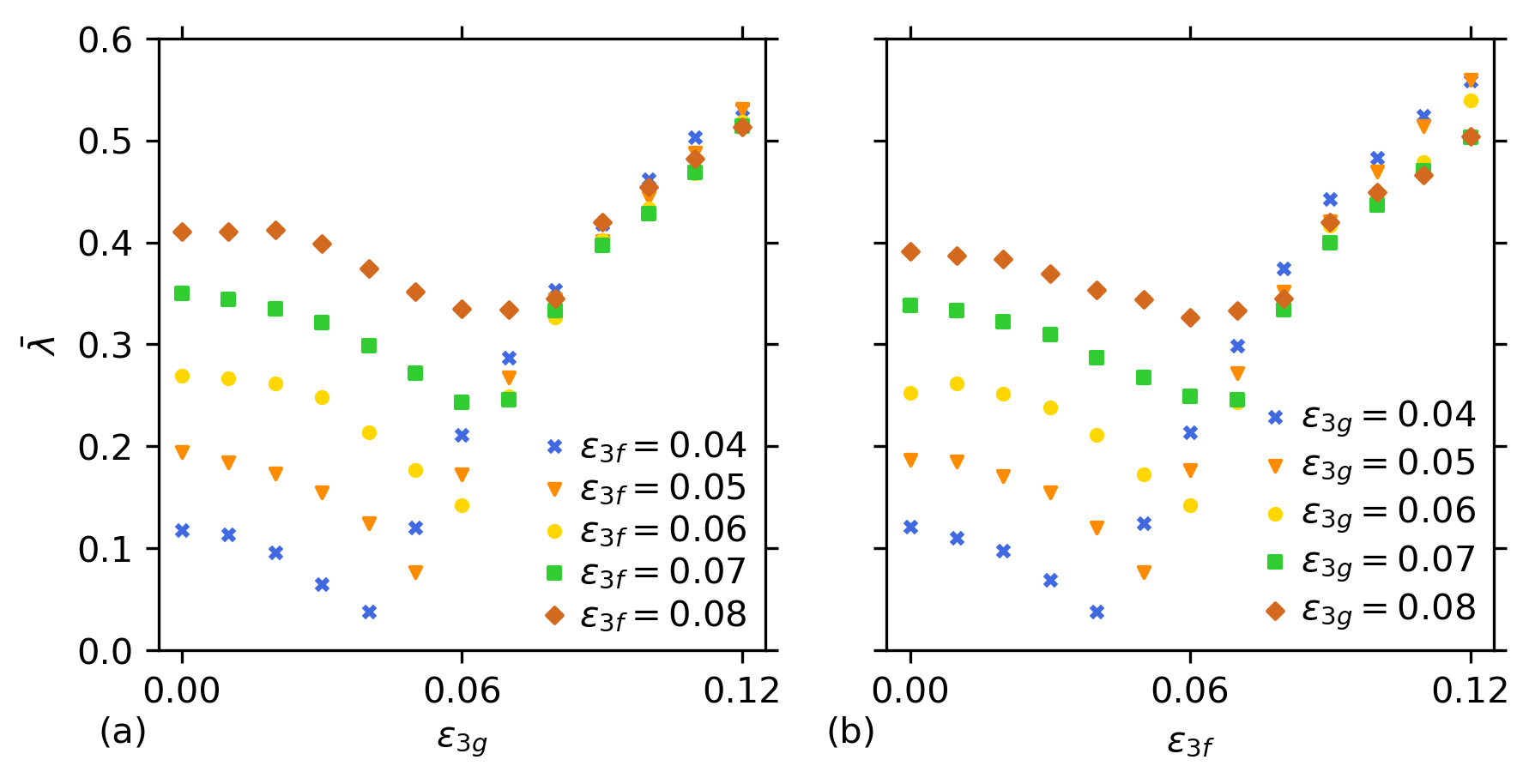}
    \caption{Average Lyapunov exponent $\bar{\lambda}$ as a function of the deformation parameter (a) $\epsilon_{3g}$ and (b) $\epsilon_{3f}$ at an orientation angle of $\theta_0=30^\circ$. $\epsilon_{3f}$ and $\epsilon_{3g}$ were left constant at values varying from $\epsilon_{const}=0.04$ to 0.08. If $\epsilon_{const}\leq0.06$, then $\bar{\lambda}$ reaches a minimum where $\epsilon_{3g}$ and $\epsilon_{3f}$ coincide.}
    \label{fig:chap2_lja3}
\end{figure}

\section{Strain as source of irregular anisotropy}
\label{chapstrain}

In this section we investigate how a distortion of the underlying (graphene) lattice induced by an uniaxial strain force will affect the mesoscopic electron dynamics (again, we focus on one valley). We illustrate the behavior on BLG with an initially trigonally warped Fermi line that is further perturbed by external strain \cite{Liu2012PhysRevB,Naumis2017Electronic,Midtvedt2016Straindisplacement}, however, our results can be generalized to other strained systems. 
After discussing the overall influence of strain on the phase-space structure, we illustrate exemplarily how increasing strain alters the details of the island structure. 

\subsection{Modeling of uniaxially strained graphene-type media}

In order to model the mechanical deformation of a hexagonal lattice  within Hooke's regime, we consider one carbon atom trigonally connected to three neighbors via the bonding vectors $\vec{\tau}_1$, $\vec{\tau}_2$ and $\vec{\tau}_3$, cf.~Fig.~\ref{chap3_strain_sketch}. It is then always possible to assume the external, uniaxial force $\vec{F}_\sigma$ to act on one of the outer atoms, and being compensated by forces $\vec{F}_\sigma/2$ acting on the two remaining outer atoms. 
Furthermore, we assume a harmonic potential $V$ for both the 
angular deformations $d\varphi$ and the binding distance deformations $dr$, 
$\vec{\tau} := r_0 \cdot [ \, \text{cos}(\varphi_0), \text{sin}(\varphi_0) \, ]^\text{T}$, (cf.~Fig.~\ref{chap3_strain_sketch}),
\begin{equation}
    V_r(dr) \; = \; \frac{D_r}{2} \, (r_0 + dr)^2 \:,
    \quad \quad
    V_\varphi(d\varphi) \; = \; \frac{D_\varphi}{2} \, (\varphi_0 + d\varphi)^2 \:.
\end{equation}
Here, $D_r$ and $D_\varphi$ are the corresponding lattice and material dependent spring constants expressing the 
proportionality 
between the applied force and the resulting deformation $d(F_\sigma)_i = -D_i \, dq_i$ with $q_i \in \{ dr, d\varphi \}$ in Hooke's regime.

In order to obtain the resulting distorted lattice corresponding to a given force $\vec{F}_\sigma$, we increase the force successively by introducing a switch-on factor $S$, starting from $S=0$.
In each step, the force was increased by $\vec{F}_\sigma/S$ and the resulting lattice distortion was computed assuming linear response, until the final value $\vec{F}_\sigma / S = \vec{F}_\sigma$ was reached, to ensure quasistatic conditions in each step.
Numerically we used 1000 steps for $S$ 
since the distorted lattice vectors $\vec{\tau}'_1$, $\vec{\tau}'_2$ and $\vec{\tau}'_3$ did not change any further even for more values of $S$.
Fig.~\ref{chap3_strain_sketch} (b) shows the decomposition of the bonding vectors distortion due to an infinitesimal increase of $F_\sigma$ in its radial and angular component.

\begin{figure}[!h]
    \centering
    \includegraphics[width=0.7\textwidth]{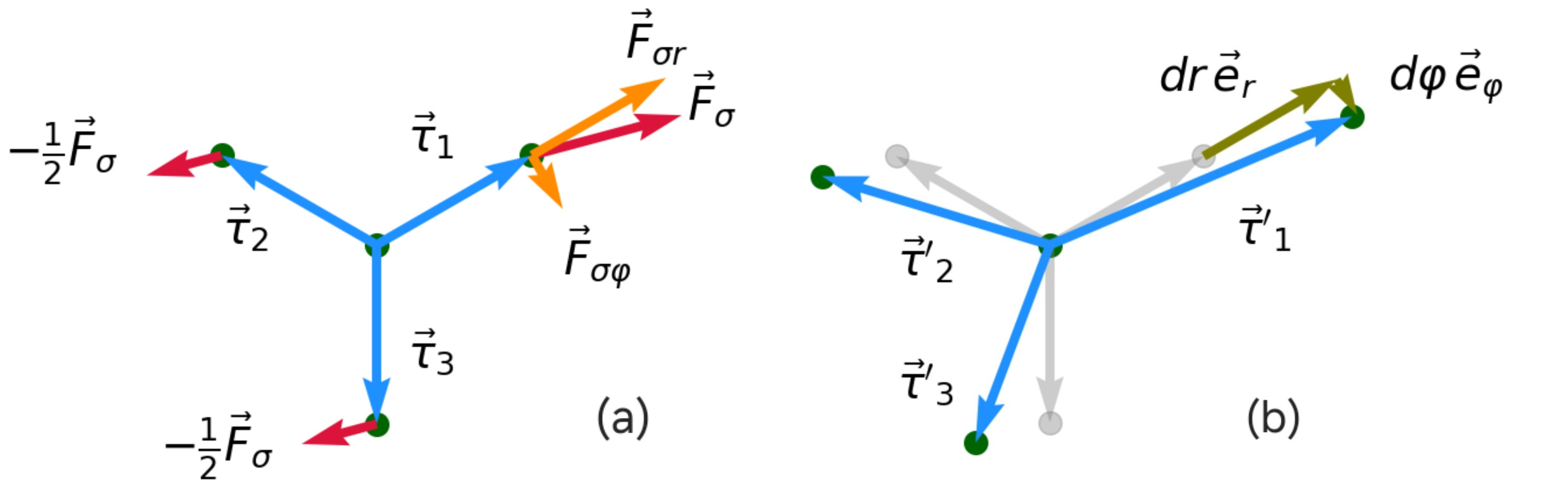}
    \caption{Schematic illustration of the bonding vectors $\vec{\tau}_1$, $\vec{\tau}_2$ and $\vec{\tau}_3$ of a hexagonal lattice unit cell.
    (a): The red vectors indicate the uniaxial force $\vec{F}_\sigma$ which acts equally on the opposite sites of the unit cell surface, therefore on the outer atoms such that the net force on the central atom vanishes.
    The orange vectors show the decomposition of $\vec{F}_\sigma$ into its radial and angular component.
    The decomposition of the corresponding distortion due to an infinitesimal increase in $\vec{F}_\sigma$ (exaggerated for illustration) is shown in (b).}
    \label{chap3_strain_sketch}
\end{figure}

Having computed the distorted lattice resulting from the mechanical strain, we now have to quantify  
its effect on the Fermi line. This is done by inserting the distorted lattice vectors $\vec{\tau}_1$, $\vec{\tau}_2$ and $\vec{\tau}_3$ into the dispersion relation of the underlying lattice (that can result, e.g., from an analytical tight-binding calculation as in single layer graphene, or can be obtained numerically). Here, we use a generic trigonally warped Fermi line inspired by BLG graphene, as an ad hoc model system with an isotropic undistorted initial state. We 
parameterize the
undistorted Fermi vector $\vec{k}_\text{Fl}$ as function of the polar angle $\lambda$ in momentum space in the parametric component form
\begin{equation}
    \vec{k}_\text{Fl}(\lambda)
    \; := \;
    \Big[ \,
    a_1 \, \text{cos}(\lambda) - a_2 \, \text{cos}(2\lambda)/\sqrt{3} \, , \,
    a_1 \, \text{sin}(\lambda) + a_2 \, \text{sin}(2\lambda)/\sqrt{3}
    \, \Big]^\text{T} \:,
    \label{Fermi_line_analytic}
\end{equation}
with $a_1$ being the radius of the underlying isotropic Fermi line. The parameter $a_2$ induces  
the trigonally warped shape, 
with $a_2$ marking the maximum deviation from the isotropic Fermi line circle reached at angles $\lambda = \pm\pi/3, \pi$. 

The strain induced deformation is assumed to be small (less than 10\%)  and can therefore be modeled using a linear stretching transformation $\mathbf{T}$ of the form
\begin{equation}
    T_{ij} \; = \;
    \delta_{ij} + (\sigma - 1) \, n_i \, n_j
    \quad \text{or} \quad
    \mathbf{T}(\vec{n}, \sigma) \; = \;
    \mathbf{1} + (\sigma - 1) \, \vec{n} \otimes \vec{n} \:.
\end{equation}
Here, $\sigma$ is the linear stretching factor in $k$-space, which is the reciprocal of the stretching factor $\sigma_l$ of the lattice in position space, $\sigma=1/\sigma_l$. The lattice stretching factor $\sigma_l$ is obtained directly from the distorted lattice vectors $\vec{\tau}_1, \vec{\tau}_2, \vec{\tau}_3$ computed above. The (uniaxial) stretching direction in real space direction $\varphi_s$ is given by the normal vector $\vec{n}$, 
$\vec{n} = [\, \text{cos}(\varphi_\text{s}), \text{sin}(\varphi_\text{s}) \,]^\text{T}$.

The strained Fermi line $\vec{k}'_\text{Fl}$ is thus given by
\begin{equation}
    \vec{k}'_\text{Fl}(\lambda) \; = \;
    \mathbf{T}(\vec{n}, s) \cdot
    \vec{k}_\text{Fl}(\lambda) \quad .
    \label{eq:FLdist}
\end{equation}

We can now model the complex dynamics of electrons in a strained system using the distorted Fermi line of Eq.~(\ref{eq:FLdist}) and investigate it in real and phase space  for various parameters $ a_1$, $a_2$, $\sigma$ and $\varphi_s$ as in the previous chapter.  The results are shown in Figs.~\ref{figstrain11} and \ref{figstrain12}.

The literature suggests that Eq.~(\ref{eq:FLdist}) is physically consistent with Fermi line shapes expected in experiments for strained BLG with
the physical parameters strain strengths $\vec{F}_s$, the ratio $D_r$/$D_\varphi$, and Fermi energy $E_F$ \cite{Roehrl2008Raman}.

To this end, we minimized the functional

\begin{equation}
    \mathcal{F}[a_1, a_2, s, \varphi_s] \; := \;
    \int_0^{2\pi} \Big| E(\vec{k'}_\text{Fl}(\lambda)) - E_\text{F} \Big| \, d\lambda
\end{equation}
where $E(\vec{k'}_\text{Fl}(\lambda))$ is the dispersion relation for distorted BLG from tight binding calculations (using the distorted bonding vectors $\vec{\tau}_i$), to find the best fit for the parameters $a_1,a_2,\sigma,\varphi_s$.

The maximum of the relative error
\begin{equation}
    \frac{\mathcal{F}}{E_\text{F}} \; := \;
    _\Delta \mathcal{F}_\text{rel}^\text{max}
    \; \approx \; 4.1 \cdot 10^{-4}
\end{equation}
occurs at high electron energies (strongly pronounced trigonal shape), at the highest considered stretching factor $\sigma = 0.94$, and when the stretching direction $\vec{n}$ was not aligned along any symmetry axis of the Fermi line. Our parametrization applies to $\sigma > 0.9$. 

Since the size of the Fermi line is irrelevant for the billiard dynamics, we can discard one of the two parameters in the analytic expression from Eq. (\ref{Fermi_line_analytic}) and set $a_1 := 1$ and 
 
$a_2 := \alpha$. 
The parameter $\alpha$  sets the shape of the Fermi line ($\alpha \rightarrow 0$: more circular, $\alpha \rightarrow 0.4$: more triangular).

\subsection{Mesoscopic electron dynamics in strained graphene based media}

In general a small uniaxial compression of the Fermi line, as mentioned above, will change the shape, size and location of the six main stability islands in the PSOS.
In addition, smaller islands may arise or disappear.
Notice that a symmetry of the cavity geometry and/or the dispersion relation (Fermi line) 
is always reflected in a symmetry in the PSOS. We shall now discuss what this implies if a onigiri-type ($C_3$) Fermi line is subject to mechanical strain. 

A compression of the Fermi line preserving one axial symmetry axis is shown in
Figures \ref{a_0.3__s_1} and \ref{a_0.3__s_0.96}. 
The three-island chain is dissolved and the islands are noticeably smaller.
The corresponding triangular orbits 
are elongated reflecting the superimposed $C_{2v}$ symmetry 
of the Fermi line.
We illustrate the effect of increasing uniaxial strain $\sigma$ by investigating the change in the structure of one of the three $k_\parallel >0$ stable islands in the PSOS by generalizing ideas of the Poincaré-Birkhoff theorem, see Fig.~\ref{figstrain13}.

We start with a weak triangular (onigiri) distortion of the Fermi line without mechanical strain in Fig.~\ref{figstrain13}a). The island has the well-known structure formed by stable trajectories revolving around the central stable fixed point. (Notice that this fixed point can be consideren to result from a Poincaré-Birkhoff -like mechanism when destroying the circular symmetry of the Fermi line resulted in a chain of three stable (elliptic) fixed point (with stable islands) with three associated unstable (hyperbolic) fixed points in between.)

 Switching on the mechanical strain, Figs.~\ref{figstrain13}b)-d), 
 results in disintegration of some of the stable trajectories in a sequence  
of new elliptic and hyperbolic fixed points that appear near the boundaries of the islands according to Poincar\'e-Birkhoff theorem. The size of the islands shrinks as the amount of chaotic motion is increased.  

\begin{figure}[!h]
    \centering
    \includegraphics[width=0.9\textwidth]{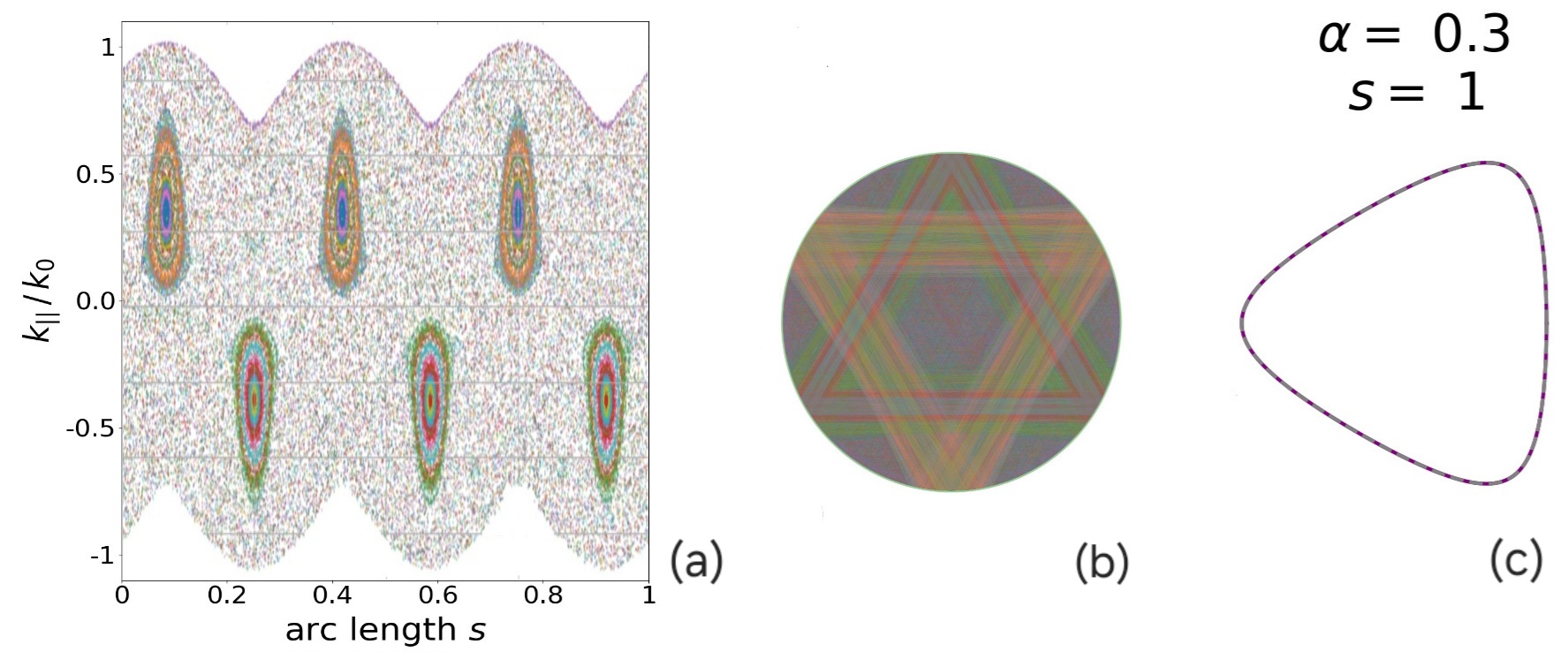}
    \caption{(a): PSOS, (b): Trajectories and (c): Fermi line for an unstrained Medium.}
    \label{a_0.3__s_1}
    \label{figstrain11}
\end{figure}

\begin{figure}[!h]
    \centering
    \includegraphics[width=0.9\textwidth]{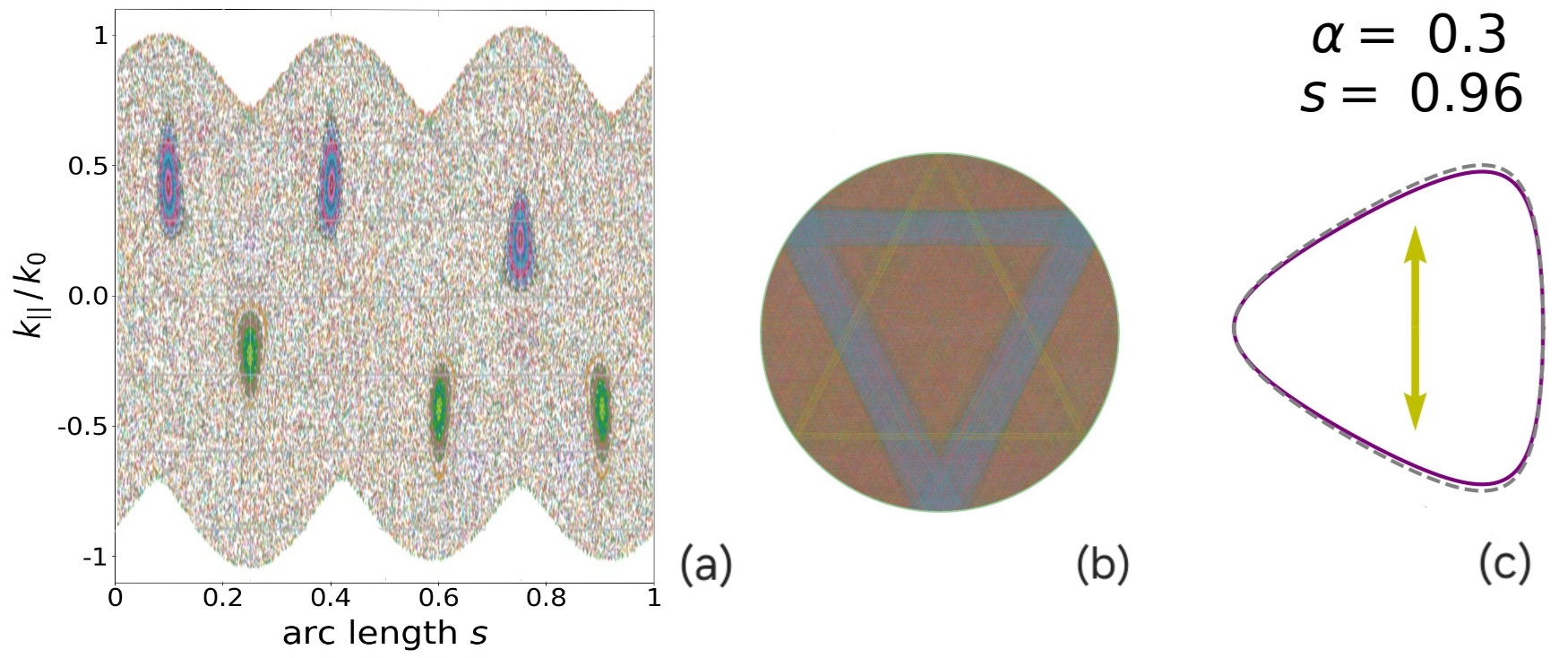}
    \caption{(a): PSOS, (b): Trajectories and (c): Fermi line for the same Fermi line as in Fig.~\ref{a_0.3__s_1} but strained with $s=0.96$ along $k_y$.}
    \label{a_0.3__s_0.96} 
    \label{figstrain12}
\end{figure}

\begin{figure}[!h]
    a)
    \begin{minipage}[t]{0.5\textwidth}
        \centering
        \includegraphics[width=.95\textwidth]{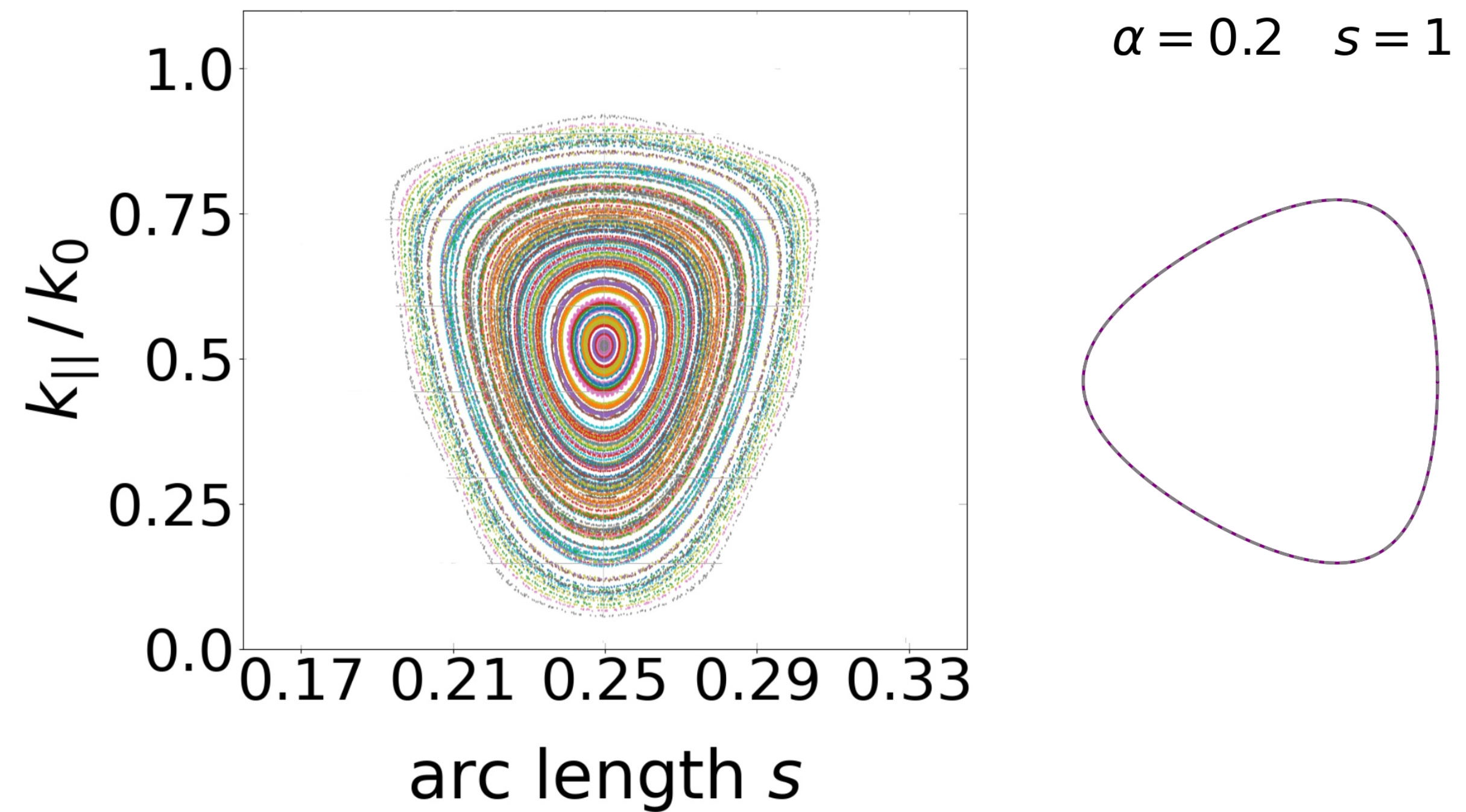}
        \label{s1}
    \end{minipage}
    b)
    \hfill
    \begin{minipage}[t]{0.5\textwidth}
        \centering
        \includegraphics[width=1\textwidth]{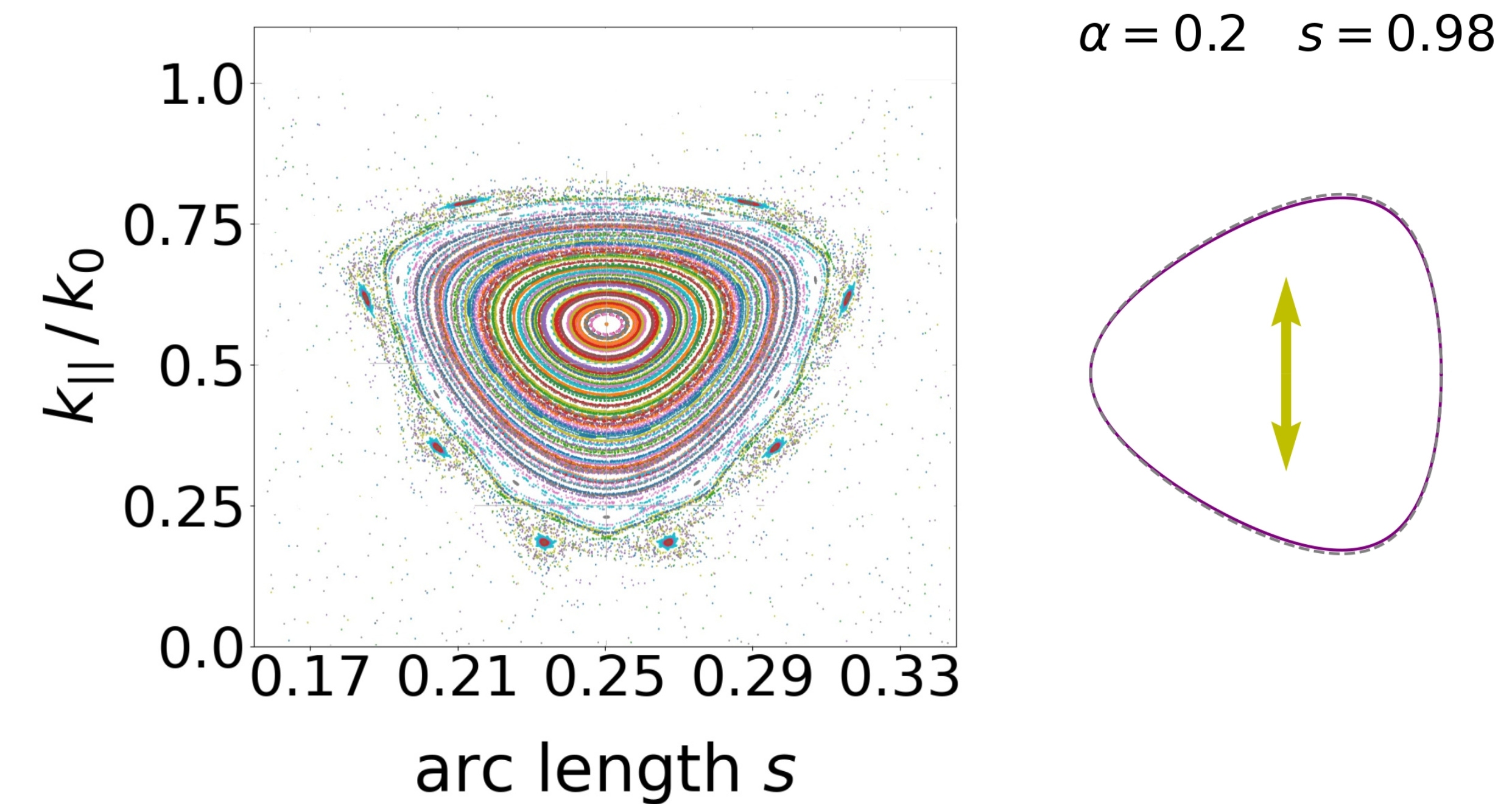}
        \label{s0.98}
    \end{minipage}

    c)
    \begin{minipage}[t]{0.5\textwidth}
        \centering
        \includegraphics[width=1\textwidth]{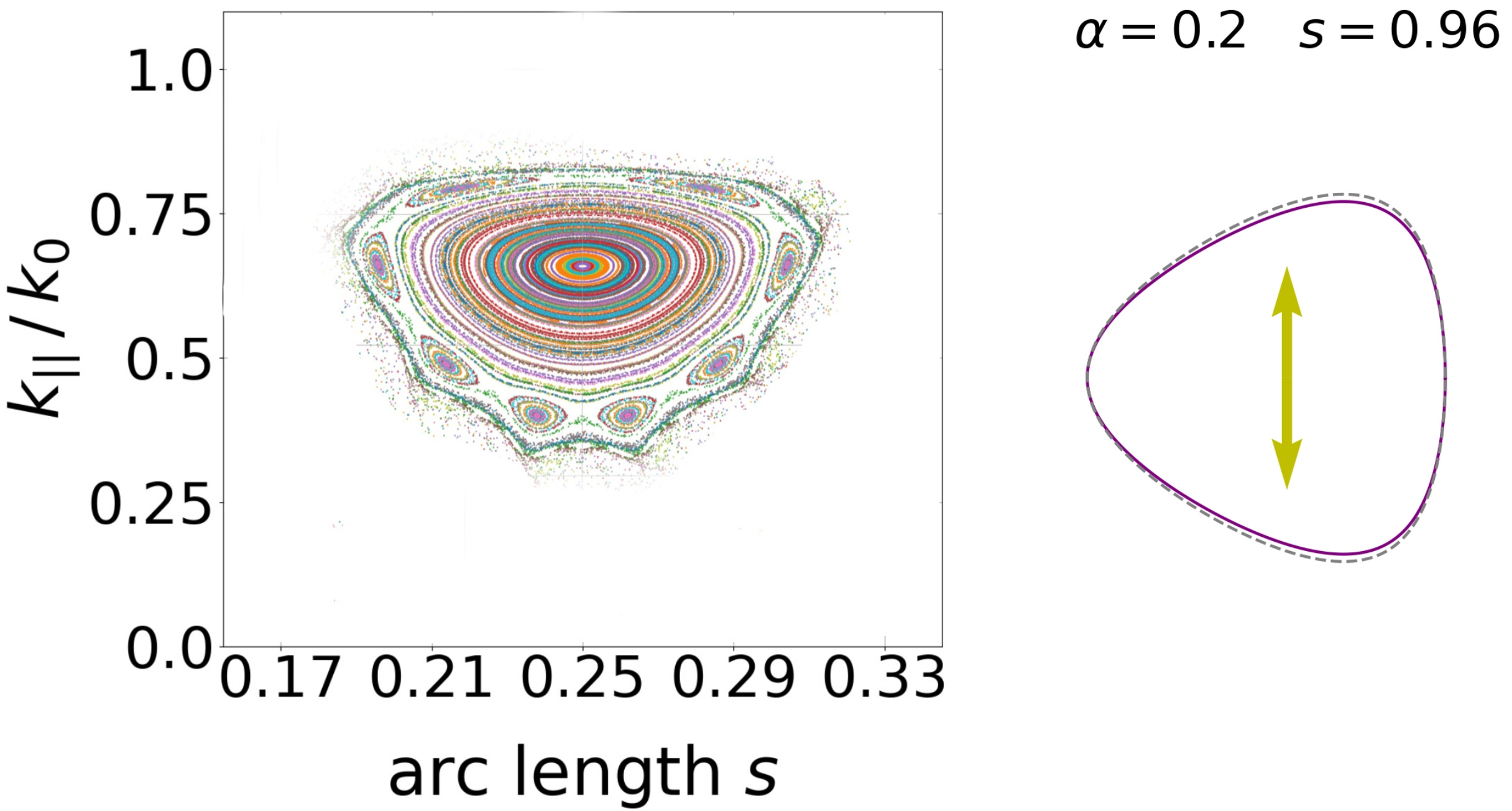}
        \label{s0.96}
    \end{minipage}
    d)
    \hfill
    \begin{minipage}[t]{0.5\textwidth}
        \centering
        \includegraphics[width=1\textwidth]{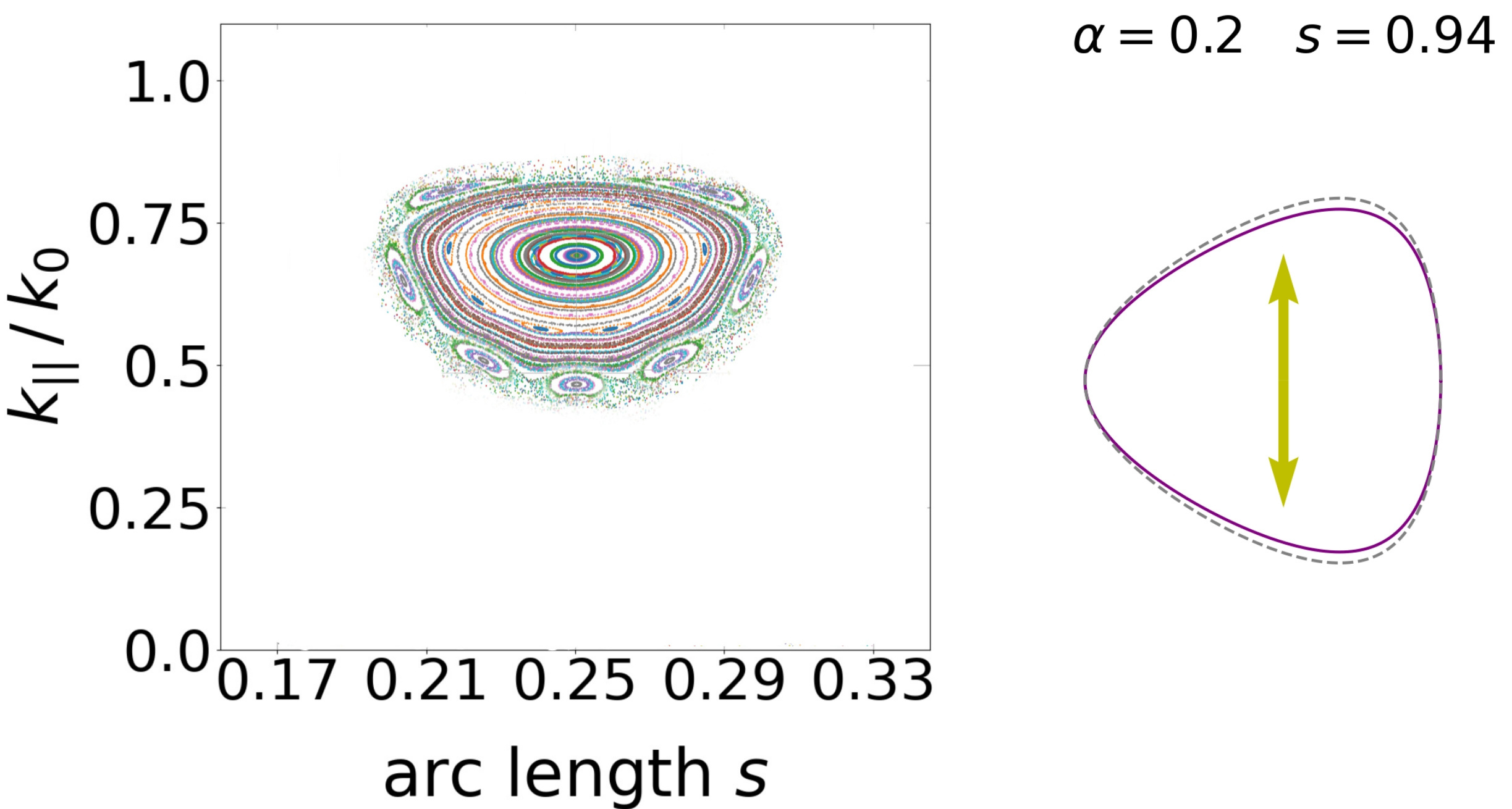}
        \label{s0.94}
    \end{minipage}
    \caption{Axial symmetric stability island with $\alpha=0.2$ for different values of the stretching factor: a) no stretch; b) $s=0.98$; c) $s=0.96$; d) $s=0.94$ along the $k_y$-axis.}
    \label{figstrain13}
\end{figure}

\section{Summary and outlook}

In this paper we have illustrated complex dynamics induced by phase space anisotropies. Using bilayer-graphene inspired model systems, we have investigated in detail the effects of deviation from an isotropic dispersion relation (circular Fermi line). We highlighted that symmetry breaking in momentum space resembles the mechanisms known from the real space counterpart where the KAM and Poincaré-Birkhoff theorems cover the transition from regular via mixed to chaotic dynamics. The advantage of operating in momentum space is, from an experimental point of view, that the dispersion relation can rather easily be manipulated all electronically for example via gate voltages, not requiring a change in the cavity geometry. Superimposing mechanical strain on the trigonally warped (onigiri) Fermi line can be used to fine-tune the complex electron dynamics in the system. 

We have compared how deviations from the circular symmetry in real and momentum space, respectively, affect the billiards dynamics. We use averaged Lyapunov exponents and find that momentum space deformation have the somewhat larger influence with the overall behavior - breaking of integrability and emergence of chaotic motion - is comparable. We point out a difference between real and momentum space deformation that is relevant for charge carriers in bilayer graphene-type systems where pseudospin or valley index is relevant. While the Poincaré SOS respects a symmetry with respect to $k_\parallel=0$ for real space deformation (clockwise and counterclockwise motions are equivalent), this is not the case for momentum space deformation where this phase space symmetry does not hold.

While we have focused on closed billiard cavities here for simplicity, there is no principal problem in describing open systems by inlcuding the generalized Fresnel-like reflection coefficients at each reflection point as in Refs.~\cite{schrepferDiracFermionOptics2021,SeemannKnHe_BLGI_PRB,Seemann_Knothe_Hentschel_2024}. Knowing the details of the dynamics of the electronic charge carriers is essential to understand their transport properties that are crucial for future applications. 

\section{Acknowledgement}
We thank Holger Kantz for helpful discussions. SG acknowledges funding by the DFG via project T1 of the Research Unit FOR 5242 (project number 449119662).

\bibliographystyle{unsrt}
\bibliography{compdynaniso}

\end{document}